\documentclass[twocolumn]{aastex62}

\newcommand\age{$\tau$}
\newcommand\mage{\tau}

\newcommand\teff{$T_\mathrm{eff}$}
\newcommand\mteff{T_\mathrm{eff}} 
\newcommand\logg{$\log g$}

\newcommand\feh{[Fe/H]}

\newcommand\mh{[M/H]}
\newcommand\afe{[$\alpha$/Fe]}
\newcommand\tife{[Ti/Fe]}
\newcommand\mtife{\mathrm{[Ti/Fe]}}

\newcommand\al{$\alpha$}

\newcommand\Qp{$Q_\mathrm{p}$}
\newcommand\Qs{$Q_\star$}

\shorttitle{Ages of Planet-hosting M dwarfs}
\shortauthors{Veyette \& Muirhead}

\begin{document}

\title{Chemo-kinematic ages of eccentric-planet-hosting M dwarf stars}

\correspondingauthor{Mark J. Veyette}
\email{mveyette@bu.edu}

\author[0000-0002-0385-2183]{Mark J. Veyette}
\affiliation{Department of Astronomy \& The Institute for Astrophysical Research, Boston University, 725 Commonwealth Ave., Boston, MA 02215, USA}

\author[0000-0002-0638-8822]{Philip S. Muirhead}
\affiliation{Department of Astronomy \& The Institute for Astrophysical Research, Boston University, 725 Commonwealth Ave., Boston, MA 02215, USA}

\begin{abstract}

M dwarf stars are exciting targets for exoplanet investigations; however, their fundamental stellar properties are difficult to measure. Perhaps the most challenging property is stellar age. Once on the main sequence, M dwarfs change imperceptibly in their temperature and luminosity, necessitating novel statistical techniques for estimating their ages. In this paper, we infer ages for known eccentric-planet-hosting M dwarfs using a combination of kinematics and $\alpha$-element-enrichment, both shown to correlate with age for Sun-like FGK stars. We calibrate our method on FGK stars in a Bayesian context. To measure $\alpha$-enrichment, we use publicly-available spectra from the CARMENES exoplanet survey and a recently developed \tife{} calibration utilizing individual \ion{Ti}{1} and \ion{Fe}{1} absorption lines in $Y$ band. Tidal effects are expected to circularize the orbits of short-period planets on short timescales; however, we find a number of mildly eccentric, close-in planets orbiting old ($\sim$8~Gyr) stars. For these systems, we use our ages to constrain the tidal dissipation parameter of the planets, \Qp{}. For two mini-Neptune planets, GJ~176~b and GJ~536~b, we find they have \Qp{} values more similar to the ice giants than the terrestrial planets in our Solar System. For GJ~436~b, we estimate an age of $8.9^{+2.3}_{-2.1}$~Gyr and constrain the \Qp{} to be $>10^5$, in good agreement with constraints from its inferred tidal heating. We find that GJ~876~d has likely undergone significant orbital evolution over its $8.4^{+2.2}_{-2.0}$~Gyr lifetime, potentially influenced by its three outer companions which orbit in a Laplace resonance.

\end{abstract}


\section{Introduction}\label{intro}

M dwarf stars are small stars, with masses between $\sim$0.1 and $\sim$0.6 $M_\sun$ and radii between $\sim$0.1 and $\sim$0.6 $R_\sun$. As a spectral class, the M type is defined by strong molecular features in their spectra, which are a consequence of their relatively cool photospheres with effective temperatures ranging between 2800 and 3800~K. M dwarf stars enable a variety of investigations into the role of stellar mass in exoplanet formation. For example, M dwarfs are known to host fewer Jupiter-mass planets than Sun-like FGK stars \citep[][]{Johnson2010}, supporting planet formation models that predict slow planetesimal growth during the protoplanetary disk phase \citep[][]{Laughlin2004}. 

Additionally, M dwarf stars enable tests of exoplanet {\it evolution} in the regime of low host-star mass. M dwarfs are known to be abundant hosts of small, short-period planets \citep{Dressing2013,Swift2013,Gaidos2014,Morton2014,Dressing2015,Gaidos2016}. Tidal interactions between a host star and its planet tend to circularize and reduce the semi-major axis of short-period orbits over time \citep{Goldreich1966}. The timescale of this evolution depends strongly on the semi-major axis of the orbit \citep{Jackson2008}. Planet-hosting M dwarfs, with their tendency to host small planets on compact orbits, are excellent targets for investigating the role of tidal migration and circularization.

Planet orbital evolution around M dwarf stars could be investigated with measurements of M dwarf ages; however, measuring ages of M dwarf stars is challenging. Once on the main-sequence, M dwarfs move imperceptibly on a Hertzsprung-Russell or color-magnitude diagram due to their low core fusion rates, taking tens of billions of years to change by a significant degree in temperature or luminosity \citep[][]{Laughlin1997,Choi2016}. Gyrochronology, the study of stellar spin-down versus age,  holds some promise for measuring M dwarf ages \citep[e.g.,][]{Meibom2015, Barnes2016}. 

However, work by \citet{Irwin2011} and \citet{Newton2016} find that field mid-M dwarfs exhibit a bimodal distribution in rotation period, similar to what is seen for Sun-like stars in young clusters \citep{Attridge1992,Barnes2003}. This indicates that either slow or fast rotation is frozen in during formation, or some rapid process takes place where stars spin down suddenly from fast to slow rotation. This could be a result of dramatic reorganizing of magnetic field topology at some specific rotation or age \citep{Garraffo2015,Garraffo2018}. If the transition is stochastic, then gyrochronology can do little to constrain the ages of young and intermediate-age M dwarfs. The prospects for applying gyrchronology after such a transition remain to be seen.

The chemical and kinematic evolution of the Galaxy provides a new way to estimate the ages of M dwarf stars. Work by \citet{Haywood2013} showed a strong correlation between stellar age, iron abundance, and \al{}-enhancement for nearby F, G and K-type dwarfs, for which ages were measured by comparing spectroscopic parameters to stellar evolution models. Work by \citet{Bensby2014} shows similarly strong correlations between \al{}-enhancement, specifically titanium enhancement (\tife{}), iron abundance, and age, over ages that span nearly the entire history of the Universe: 1.5 to 13.5 Gyr. The relation between \al{}-enhancement and stellar age is the result of early ISM enrichment of \al{} elements by core-collapse supernovae and delayed enrichment of iron by Type Ia supernovae. The delayed enrichment of iron causes \afe{} to decrease and \feh{} to increase over time. This trend has been confirmed by numerous studies of solar-neighborhood FGK stars \citep{Nissen2015,Spina2016,Buder2018} and red giant stars \citep{Martig2015,Hawkins2016,Feuillet2016,Feuillet2018}. Recently, \citet{Bedell2018} showed that when restricting the stellar sample to only solar twins (stars with similar temperature, surface gravity, and overall metallicity to the Sun), there is an exceptionally tight relation between stellar age and the abundance of alpha elements, including titanium. These age-abundance relations provide a path to statistically {\it measure} stellar ages, and should apply just as well to nearby M dwarfs as they does to nearby F, G and K-type stars.

M dwarfs' cooler photospheres allow molecules to form throughout their atmospheres. Opacity from these molecules contribute millions of absorption lines that blanket an M dwarf's optical and NIR spectrum. Difficulties in modeling cool stellar atmospheres and the millions of molecular transitions occurring in them have so far prohibited the detailed chemical analysis of M dwarfs. Empirically calibrated, model-independent methods to measure M dwarf metallicities provide a way around these issues \citep{Bonfils2005,Johnson2009,Rojas2010,Mann2013a}. However, these methods are indirect tracers of metallicity, relying on astrophysical abundance correlations \citep{Veyette2016b}, and are limited to measuring overall metallicity. \citet{Veyette2017} presented a new physically motivated and empirically calibrated method to measure the effective temperature, iron abundance, and titanium enhancement of an M dwarf from its high-resolution $Y$-band spectrum around 1~$\micron$. With the ability to measure \tife{} of M dwarfs, we can now apply the well-studied \tife{}-age relation to estimate ages of M dwarfs.

In this paper, we estimate ages for eccentric-planet-hosting M dwarf stars by combining galactic kinematics with titanium-enhancement. In Section~\ref{sec:sample} we describe how the sample of planet-hosts was chosen and the high-resolution NIR spectra used in this work. In Section~\ref{sec:analysis} we describe how we measured \tife{} of these M dwarfs from their high-resolution $Y$-band spectra and how we used a sample of FGK stars with measured \tife{} and ages to calibrate an empirical, probabilistic \tife{}-age-relation. In Section~\ref{sec:results} we combine our \tife{}-age relation with a kinematic prior to estimate ages for our sample of planet-hosting M dwarfs. In Section~\ref{sec:discussion} we use our ages to explore the tidal evolution of the planets and constrain their tidal $Q$. Finally, we summarize this work in Section~\ref{sec:summary}.

\section{Sample}\label{sec:sample}

Calar Alto high-Resolution search for M dwarfs with Exo-earths with Near-infrared and optical Échelle Spectrographs (CARMENES) is a high-resolution optical and NIR spectroscopic survey to search for rocky planets in the habitable zones of nearby M dwarfs \citep{Quirrenbach2014}. The CARMENES spectrograph covers 0.5 to 1.7 $\micron$ at a resolution of 94,600 in the optical and 80,400 in the NIR. \citet[][hereafter R17]{Reiners2017} published one representative CARMENES spectrum for each of 324 M dwarfs in the survey.

We downloaded the NIR spectra for all 324 CARMENES GTO targets from the CARMENES GTO Data Archive \citep{Caballero2016}\footnote{\url{http://carmenes.cab.inta-csic.es/}}. Many of the spectra exhibit large, spurious features that are likely a result of the automatic flat-relative extraction pipeline used by CARMENES \citep{Zechmeister2014}. We checked each spectrum by eye in the $Y$-band region and exclude from further analysis any spectrum that contains either large spikes spanning over 100 pixels that are present in multiple orders or large, sharp variations in the continuum that make it impossible to consistently assess the pseudo-continuum across the full $Y$ band. Roughly half the spectra did not meet our quality cuts. We also exclude stars with projected rotational velocity $v \sin i > 12$ km s$^{-1}$, corresponding to the resolution of the NIRSPEC spectra used to calibrate the \citet{Veyette2017} method.

We cross-matched the stars the passed our quality cuts and that had masses $> 0.2 M_\sun$ with the NASA Exoplanet Archive\footnote{\url{https://exoplanetarchive.ipac.caltech.edu/}}. We found 11 M dwarfs that host known exoplanets: GJ~176~b \citep{Forveille2009}, GJ~179~b \citep{Howard2010}, GJ~436~b \citep{Butler2004}, GJ~536~b \citep{SuarezMascareno2017a}, GJ~581~b,~c,~e \citep{Bonfils2005b,Udry2007,Mayor2009}, HD~147379~b \citep[GJ~617~A,][]{Reiners2018}, GJ~625~b \citep{SuarezMascareno2017b}, Wolf~1061~b,~c,~d \citep[GJ~628,][]{Wright2016}, GJ~649~b \citep{Johnson2010b}, GJ~849~b \citep{Butler2006}, and GJ~876~b,~c,~d,~e \citep{Marcy1998,Marcy2001,Rivera2005,Rivera2010}. Table~\ref{tbl:planets} lists the exoplanets analyzed in this study and their relevant parameters.

\begin{deluxetable*}{lDDDc}
\caption{Planet-hosting M dwarf Exoplanet Parameters \label{tbl:planets}}
\tablehead{
\colhead{Planet} & \multicolumn2c{$M_p\sin i$ [$M_\earth$]} & \multicolumn2c{$a$ [AU]} & \multicolumn2c{$e$} & \colhead{Ref.}}
\decimals
\startdata
GJ 176 b & $9.06^{+1.54}_{-0.7}$ & $0.066^{+0.001}_{-0.001}$ & $0.148^{+0.249}_{-0.036}$ & 1 \\
GJ 179 b & $260.61^{+22.25}_{-22.25}$ & $2.41^{+0.04}_{-0.04}$ & $0.21^{+0.08}_{-0.08}$ & 2 \\
GJ 436 b & $21.36^{+0.2}_{-0.21}$ & $0.028^{+0.001}_{-0.001}$ & $0.152^{+0.009}_{-0.008}$ & 1 \\
GJ 536 b & $6.52^{+0.69}_{-0.4}$ & $0.067^{+0.001}_{-0.001}$ & $0.119^{+0.125}_{-0.032}$ & 1 \\
GJ 581 b & $15.2^{+0.22}_{-0.27}$ & $0.041^{+0.001}_{-0.001}$ & $0.022^{+0.027}_{-0.005}$ & 1 \\
GJ 581 c & $5.652^{+0.386}_{-0.239}$ & $0.074^{+0.001}_{-0.001}$ & $0.087^{+0.15}_{-0.016}$ & 1 \\
GJ 581 e & $1.657^{+0.24}_{-0.161}$ & $0.029^{+0.001}_{-0.001}$ & $0.125^{+0.078}_{-0.015}$ & 1 \\
GJ 617A b & $24.7^{+1.8}_{-2.4}$ & $0.3193^{+0.0002}_{-0.0002}$ & $0.01^{+0.12}_{-0.01}$ & 3 \\
GJ 625 b & $2.82^{+0.51}_{-0.51}$ & $0.078361^{+4.4e-05}_{-4.6e-05}$ & $0.13^{+0.12}_{-0.09}$ & 4 \\
GJ 628 b & $1.91^{+0.26}_{-0.25}$ & $0.0375^{+0.0012}_{-0.0013}$ & $0.15^{+0.13}_{-0.1}$ & 5 \\
GJ 628 c & $3.41^{+0.43}_{-0.41}$ & $0.089^{+0.0029}_{-0.0031}$ & $0.11^{+0.1}_{-0.07}$ & 5 \\
GJ 628 d & $7.7^{+1.12}_{-1.06}$ & $0.47^{+0.015}_{-0.017}$ & $0.55^{+0.08}_{-0.09}$ & 5 \\
GJ 649 b & $104.244^{+10.17}_{-10.17}$ & $1.135^{+0.035}_{-0.035}$ & $0.3^{+0.08}_{-0.08}$ & 6 \\
GJ 849 b & $289.21$ & $2.32$ & $0.05^{+0.03}_{-0.03}$ & 7 \\
GJ 876 b & $760.9^{+1.0}_{-1.0}$ & $0.214^{+0.001}_{-0.001}$ & $0.027^{+0.002}_{-0.002}$ & 1 \\
GJ 876 c & $241.5^{+0.7}_{-0.6}$ & $0.134^{+0.001}_{-0.001}$ & $0.25^{+0.001}_{-0.002}$ & 1 \\
GJ 876 d & $6.91^{+0.22}_{-0.27}$ & $0.021^{+0.001}_{-0.001}$ & $0.082^{+0.043}_{-0.025}$ & 1 \\
GJ 876 e & $15.43^{+1.29}_{-1.27}$ & $0.345^{+0.001}_{-0.002}$ & $0.04^{+0.021}_{-0.004}$ & 1 \\
\enddata
\tablecomments{$M_p$ is used when $i$ is known. References: (1) \citet{Trifonov2018}, (2) \citet{Howard2010}, (3) \citet{Reiners2018}, (4) \citet{SuarezMascareno2017b}, (5) \citet{Astudillo-Defru2017}, (6) \citet{Johnson2010}, (7) \citet{Bonfils2013}}
\end{deluxetable*}

\section{Analysis}\label{sec:analysis}

\subsection{Measuring \tife{}}

We employed the method developed by \citet{Veyette2017} to measure the \teff{}, \feh{}, and \tife{} of the M dwarfs in our sample from the $Y$-band region of their high-resolution CARMENES spectra. The method utilizes strong, relatively isolated Fe and Ti lines in $Y$ band to directly estimate Fe and Ti abundances. The method is physically motivated, using a custom grid of \texttt{PHOENIX} BT-Settl models \citep{Allard2012a,Baraffe2015,Allard2016} to provide the nonlinear relations for how M dwarf spectra change as a function of temperature and composition. It is also empirically calibrated by using observations of widely separated FGK and M type binary stars to derive corrections to the model relations, ensuring agreement between abundance analyses of solar-type stars and M dwarfs.

The \citet{Veyette2017} method was originally calibrated on Keck/NIRSPEC spectra at a resolution of 25000. Due to severe blending with neighboring molecular lines, the calibration is only valid when applied to spectra at the same resolution. We took the following steps to prepare the CARMENES spectra and closely match the format of the NIRSPEC spectra used in the original calibration. First, we masked out pixels 300-370 in the 3rd and 4th orders of the NIR spectra. Most spectra had broad peaks at these pixel locations which we assume are an artifact of the reduction process. Next to remove a number of large narrow spikes that appear at random pixel locations throughout many of the spectra, we masked out pixels with flux values that were more than five median absolute deviations greater than the median flux value of the 200 surrounding pixels. We then interpolated the spectra to a finer grid with uniform log-spacing in wavelength and convolved them down to a resolution of 25000. Finally, to remove edge effects from the discrete convolution, we masked out pixels within $\pm$2.5 times the convolution kernel FWHM of the edge of each order or the chip gap at the 2040th pixel location.

\begin{deluxetable*}{lDDDDDDD}
\caption{Planet-hosting M dwarf Stellar Parameters \label{tbl:ages}}
\tablehead{\colhead{Name} & \multicolumn2c{U [km/s]} & \multicolumn2c{V [km/s]} & \multicolumn2c{W [km/s]} & \multicolumn2c{\teff{} [K]} & \multicolumn2c{\feh{}} & \multicolumn2c{\tife{}} & \multicolumn2c{Age [Gyr]}}
\decimals
\startdata
GJ 176 & $-22.6$ & $-56.7$ & $-14.8$ & $3538$ & $+0.05$ & $+0.04$ & $8.8^{+2.5}_{-2.8}$ \\
GJ 179 & $+13.3$ & $-17.2$ & $+0.9$ & $3350$ & $+0.12$ & $+0.06$ & $4.6^{+3.5}_{-2.4}$ \\
GJ 436 & $+52.0$ & $-19.2$ & $+19.5$ & $3466$ & $-0.08$ & $+0.07$ & $8.9^{+2.3}_{-2.1}$ \\
GJ 536 & $-54.6$ & $+2.3$ & $+2.9$ & $3653$ & $-0.12$ & $+0.06$ & $6.9^{+2.5}_{-2.3}$ \\
GJ 581 & $-25.0$ & $-25.4$ & $+11.6$ & $3377$ & $+0.06$ & $+0.04$ & $6.6^{+2.9}_{-2.5}$ \\
GJ 617A & $-9.9$ & $-30.1$ & $+4.0$ & $3966$ & $+0.13$ & $-0.00$ & $5.1^{+3.2}_{-2.4}$ \\
GJ 625 & $+7.8$ & $-2.6$ & $-17.6$ & $3433$ & $-0.37$ & $+0.12$ & $7.0^{+2.7}_{-4.1}$ \tablenotemark{a} \\
GJ 628 & $-13.0$ & $-21.1$ & $-20.6$ & $3456$ & $-0.25$ & $-0.02$ & $4.3^{+3.1}_{-2.0}$ \\
GJ 649 & $+21.3$ & $-14.3$ & $+1.2$ & $3595$ & $+0.03$ & $-0.02$ & $4.5^{+3.0}_{-2.0}$ \\
GJ 849 & $-44.6$ & $-17.6$ & $-17.6$ & $3469$ & $+0.28$ & $-0.01$ & $4.9^{+3.0}_{-2.1}$ \\
GJ 876 & $+1.3$ & $-2.2$ & $-49.9$ & $3295$ & $+0.18$ & $+0.02$ & $8.4^{+2.2}_{-2.0}$ \\
\enddata
\tablecomments{Ages are posterior medians and $\pm$1 $\sigma$ values corresponding to 16th and 84th percentiles.}
\tablenotetext{a}{See discussion in Section~\ref{sec:GJ625}}
\end{deluxetable*}

We followed the same procedure outlined in \citet{Veyette2017} to correct the shape of each order and set the pseudo-continuum level. We excluded the 1.05343--1.05360~$\micron$ Fe line and the 1.07285--1.07300~$\micron$ Ti line from our analysis as they fall too close to the chip gap and were masked out in some spectra. We also changed the FeH index defined in \citet{Veyette2017} to be the ratio of the flux in the 0.988--0.9895$\micron$  and 0.990--0.992~$\micron$ regions. The original definition of the FeH index covered the chip gap and a transition from one order to the next. We used the calibration sample from \citet{Veyette2017} to recalculate the empirical corrections for this modified feature list. The accuracy of the calibration is similar to that achieved with the original feature list. The RMSE of the inferred parameters of our calibration sample are 56~K in \teff{}, 0.12~dex in \feh{}, and 0.05~dex in \tife{}. Using this new calibration and the cleaned CARMENES spectra, we measure the \teff{}, \feh{}, and \tife{} of the M dwarfs in our planet-hosting sample. The results are listed in Table~\ref{tbl:ages}.

\subsection{A Bayesian estimate of stellar ages}
Starting with Bayes' Theorem, the posterior probability distribution of a star's age, \age{}, given its titanium enhancement, \tife{}, and our prior information, $I$, can be written as
\begin{equation} \label{eq:bayes}
    p\left( \mage{} \left| \mtife{}, I \right. \right) \propto
    p\left( \mage{} \left| I \right. \right)
    p\left( \mtife{} \left| \mage{}, I \right. \right).
\end{equation}
Here, the prior information includes three propositions: (1) a prior probability distribution for \age{} based on previous information, which in this case, will be the star's peculiar velocities, (2) a model for the likelihood of a given \tife{} measurement as a function of age, and (3) the measurement uncertainty in \tife{} can be described by a Gaussian with standard deviation $\sigma_\mtife{}$ = 0.05 dex.

\subsubsection{FGK calibration sample}
In the following sections we describe a data-driven approach to estimate the kinematic prior and \tife{} likelihood. For this approach, we require an unbiased sample of stars with known age, kinematics, and \tife{}. The \citet{Veyette2017} method to measure \tife{} was calibrated to match the \citet[][hereafter B16]{Brewer2016a} catalog of detailed abundances for 1,617 FGK stars. Therefore, to ensure consistency and reduce systematic errors, we used this same catalog to develop our kinematic--\tife{}--age model. To estimate ages of the B16 stars, we used the \texttt{isochrones} package \citep{Morton2015} with the MIST stellar evolution models \citep{Dotter2016,Choi2016}. We describe this process in detail in Appendix~\ref{fgkages}.

\begin{figure}
\centering
\includegraphics[width=\linewidth]{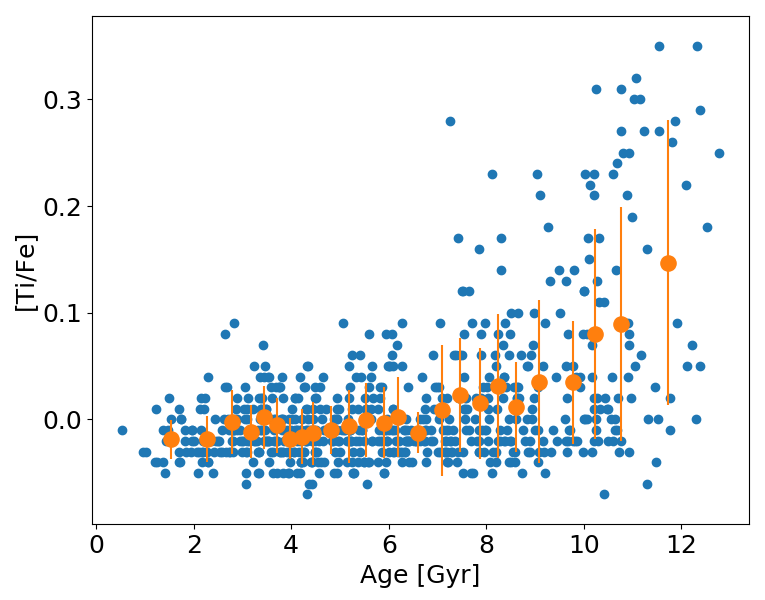}
\caption{\citet{Brewer2016a} \tife{} measurements of solar neighborhood FGK stars versus our stellar ages. Large orange circles denote the mean \tife{} in 25 age bins spaced so that each bin contains approximately the same number of stars. Error bars indicate the standard deviation in each bin.\label{fig:age_vs_tife}}
\end{figure}

B16 fit for and removed from their abundance estimates any systematic trends with temperature; however, this trend was only assessed over a limited range of \teff{} and \logg{}. We found that systematic trends in \tife{} still existed for stars with \teff{} $>$ 6100 K and \logg{} $<$ 3.6 and so we excluded those stars from further analysis. We also excluded stars with best fit $A_V$ values $>$ 0.1. All stars in this sample are solar neighborhood stars and we do not expect significant extinction. These cuts, combined with the initial requirement that stars have a parallax measurement available in the literature and convergence criteria as described in Appendix~\ref{fgkages}, leaves 672 FGK stars for which we have reliable \tife{} and age estimates. Figure~\ref{fig:age_vs_tife} shows the general trend of increasing \tife{} with increasing age.

Of these 672 stars, 658 have radial velocities and full 5-parameter astrometric solutions in Gaia DR2 \citep{GaiaDR2,Lindegren2018}, which we used to calculate the $U$, $V$, and $W$ peculiar velocities of each star (calculations based on code adapted from \citealt{Rodriguez2016}).

\subsubsection{Kinematic Prior}\label{sec:prior}

\citet[][hereafter A18]{AlmeidaFernandes2018} introduced a method for estimating the age of a star based on its peculiar velocities alone. They modeled the components of the velocity ellipsoid of field stars as Gaussian distributions with age-dependant dispersion. They used the Geneva-Copenhagen Survey \citep[GCS,][]{Nordstrom2004,Casagrande2011} to fit for the dispersion of these distributions as functions of age as well as the $V$ component of the Solar motion, $V'_\sun$, and the vertex deviation, $\ell_v$. Evaluating the product of the three distributions at a given age produces the likelihood function for measured $U$, $V$, and $W$ velocities. We employ a prior probability distribution for a star's age based on the posterior probability distribution given by Equation~10 of A18 (method $UVW$),
\begin{equation}\label{eq:kinematic_age_post}
    p\left( \mage{} \left| U,V,W \right.\right) \propto \prod_{i=1,2,3} \frac{1}{\sqrt{2\pi}\sigma_i(\mage{})} \exp{\left(-\frac{v_i^2}{2\sigma_i(\mage{})^2}\right)},
\end{equation}
where $v_1$, $v_2$, and $v_3$ are the star's velocities in terms of the components of the velocity ellipsoid as defined by Equations~4a--c of A18 and $\sigma_1(\mage{})$, $\sigma_2(\mage{})$, and $\sigma_3(\mage{})$ are power laws with parameters from Table~1 of A18.

\begin{figure}
\centering
\includegraphics[width=\linewidth]{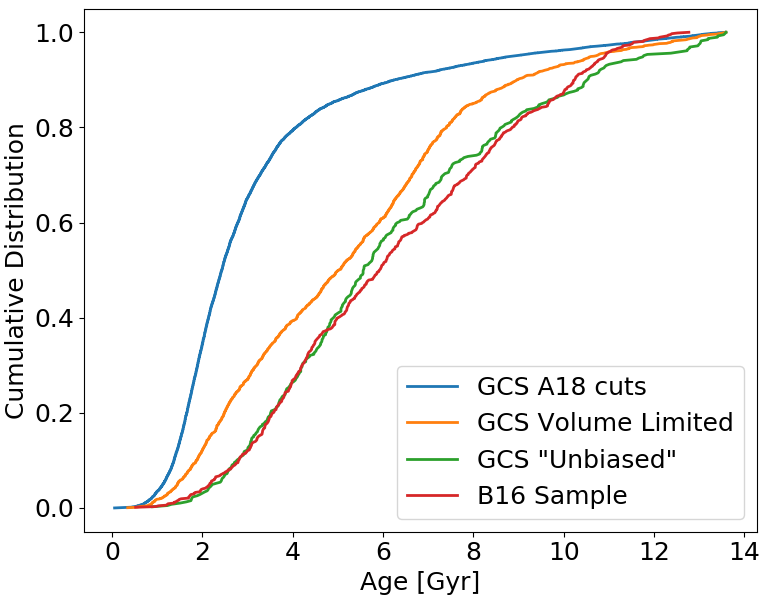}
\caption{Cumulative distributions of ages in the GCS sample used by A18, a volume-limited sample of the GCS, an ``unbiased'' sample of the GCS, and our sample of B16 stars. See Section~\ref{sec:prior} for more information on each sample. \label{fig:age_cdfs}}
\end{figure}

A18 made various cuts to the \citet{Casagrande2011} GCS catalog to ensure high-quality kinematic data and age estimates. The age distribution of their subsample is similar to the full, magnitude-limited GCS which is known to be biased toward bright F-type stars \citep{Nordstrom2004}. Since the main-sequence lifetime of a 1.1~$M_\sun$ star is roughly half the age of the Universe \citep{Choi2016}, using the full GCS significantly biases kinematic ages estimates toward younger ages. In figure~\ref{fig:age_cdfs} we show age cumulative distributions for the A18 sample compared to a volume-limited sample of the GCS ($d<40$~pc). The volume-limited sample is significantly shifted toward older ages.

This volume-limited sample still contains a number of F dwarfs with main-sequence lifetimes much shorter than the age of the Universe, which biases the sample against older stars. In an attempt to create a sample of stars that better matches to true age-distribution of low-mass stars in the solar neighborhood, we further restrict the volume-limited sample to stars with $0.9 M_\sun < M_\star < 1 M_\sun$. This restricts the sample to mainly G dwarfs whose lifetimes are not significantly shorter than the age of the Universe and for which the GCS is mostly complete out to 40~pc. The lower limit in mass excludes stars for which the ages are not well constrained. Figure~\ref{fig:age_cdfs} also shows the age distributions for this ``unbiased'' sample of the GCS and our sample of B16 stars. Although, we note that it is extremely difficult to assemble a truly unbiased and complete sample of stars. The B16 sample is comprised of stars originally observed as part of the California Planet Survey \citep{Howard2010}, an RV exoplanet survey. As such, it is biased against stars with excessive velocity jitter and faint stars. Our \logg{} and \teff{} cuts along with our requirement that each star have Tycho-2 and 2MASS magnitudes as described in Appendix~\ref{fgkages}  introduce additional biases. Overall, however, these biases do not result in any significant age bias for our final sample as evidenced by the similarity between the age distributions for our B16 sample and the volume-limited ``unbiased'' GCS sample. This is largely a result of the fact that both samples have been limited to solar-neighborhood Sun-like stars. As these stars have lifetimes on the order of the age of the Universe, their age distribution should be very similar to the age distribution of solar-neighborhood M dwarfs.

For consistency, we used our sample of B16 stars to recalibrate the A18 kinematic likelihood. We find the following best fit relations for the vertex deviation, V component of the Solar motion, and dispersion in the three components of the velocity ellipsoid as functions of time.
\begin{eqnarray}\label{eq:afrp_refit1}
    \ell_v &=& 0.406 e^{-0.163\mage{}} \\
    V'_\sun &=& 0.314 \mage{}^2 - 1.28 \mage{} + 17.0 \\
    \sigma_1 &=& 13.6 \mage{}^{0.484} \\
    \sigma_2 &=& 7.32 \mage{}^{0.493} \\
    \sigma_3 &=& 4.20 \mage{}^{0.703}\label{eq:afrp_refit2},
\end{eqnarray}
where all velocities are in km~s$^{-1}$. We also calculate the U and W components of the solar motion to be 9.33~km~s$^{-1}$ and 7.95~km~s$^{-1}$, respectively.

We used proper motions and parallaxes from \citet{Gaidos2014} along with radial velocities from R17 to calculate the $U$, $V$, and $W$ peculiar velocities for each M dwarf in our exoplanet-host sample. The velocities for each star are list in Table~\ref{tbl:ages}. We used these velocities and Equations~\ref{eq:afrp_refit1}--\ref{eq:afrp_refit2} to calculate the posterior probability distributions of the kinematic ages given by Equation~\ref{eq:kinematic_age_post}. We used these posteriors as the prior probability in Equation~\ref{eq:kinematic_age_post}~\ref{eq:bayes}.

\subsubsection{A data-driven \tife{} likelihood}\label{sec:likelihood}

Following the approach of \citet{AlmeidaFernandes2018} for constructing a data-driven likelihood function, we used a sample of FGK stars with measured \tife{} and stellar ages to calculate the likelihood of a star's measured \tife{} given an assumed age. 

\begin{figure}
\centering
\includegraphics[width=\linewidth]{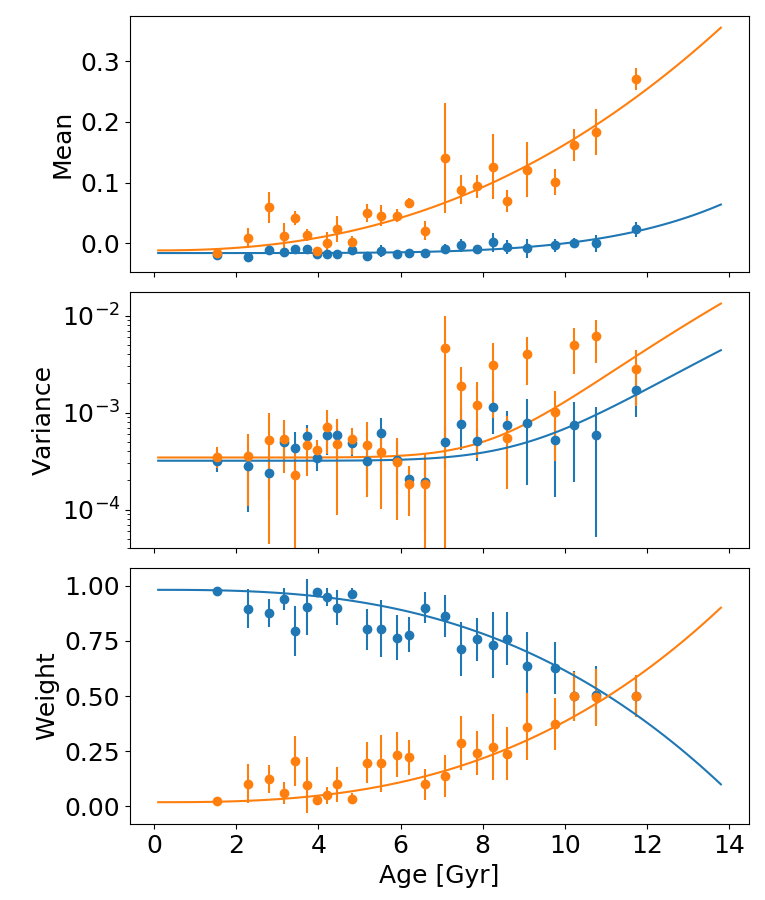}
\caption{Means, variances, and weights of the two components (blue and orange) of our Gaussian mixture model as functions of age. Errors are from bootstrap resampling within each age bin. Solid lines are fits to the model parameters based on Equation~\ref{eq:gmm_param_fits} with best fit parameters from Table~\ref{tbl:gmm_param_fits}.\label{fig:gmm_param_fits}}
\end{figure}

\begin{deluxetable}{clDDD}
\caption{Best fit constants for Equation~\ref{eq:gmm_param_fits} \label{tbl:gmm_param_fits}}
\tablehead{
\colhead{} & \colhead{} & \multicolumn2c{$a$} & \multicolumn2c{$b$} & \multicolumn2c{$c$}
}
\decimals
\startdata
Comp. 1 & mean & $-$0.0160 & 1.65$\times10^{-7}$ & 4.99 \\
& variance & 3.19$\times 10^{-4}$ & 1.12$\times 10^{-11}$ & 7.51 \\
& weight & 0.982 & $-$6.78$\times 10^{-4}$ & 2.73 \\
Comp. 2 & mean & $-$0.0116 & 8.55$\times 10^{-4}$ & 2.31 \\
& variance & 3.44$\times 10^{-4}$ & 7.00$\times 10^{-12}$ & 8.13 \\
& weight & 0.018 & 6.78$\times 10^{-4}$ & 2.73 \\
\enddata
\end{deluxetable}

We first divided our FGK sample into 25 age bins spaced so that each bin contains roughly an equal number of stars ($\sim$27 per bin). Motivated by the existence of two chemically distinct populations in the solar neighborhood \citep[e.g.,][]{Fuhrmann1998}, we modeled the \tife{} distribution within a bin as Gaussian mixture model with two components. We fit for the means, variances, and weights of each component via expectation-maximization\footnote{Specifically, we used the Scikit-learn \texttt{GaussianMixture} package \citep{scikit-learn}.}. We assessed the uncertainty in the mixture model parameters by bootstrap resampling the \tife{} distribution within an age bin and refitting the mixture model, repeating this 10000 times. In order to create a continuous likelihood as a function of age, we fit an offset power law of the form
\begin{equation}\label{eq:gmm_param_fits}
    \theta_i(\mage{}) = a+b\mage{}^c
\end{equation}
to each mixture model parameter, $\theta_i$. Here, $\mage{}$ is the average age of the stars in the bin. We determined the best fit parameters via $\chi^2$ minimization and list them in Table~\ref{tbl:gmm_param_fits}. Figure~\ref{fig:gmm_param_fits} shows our mixture model parameters as a function of age, their uncertainties, and our power law fits. Note that since the weights of the two components must sum to unity, we only fit to the weights of one component. 

\begin{figure*}
\centering
\includegraphics[width=\linewidth]{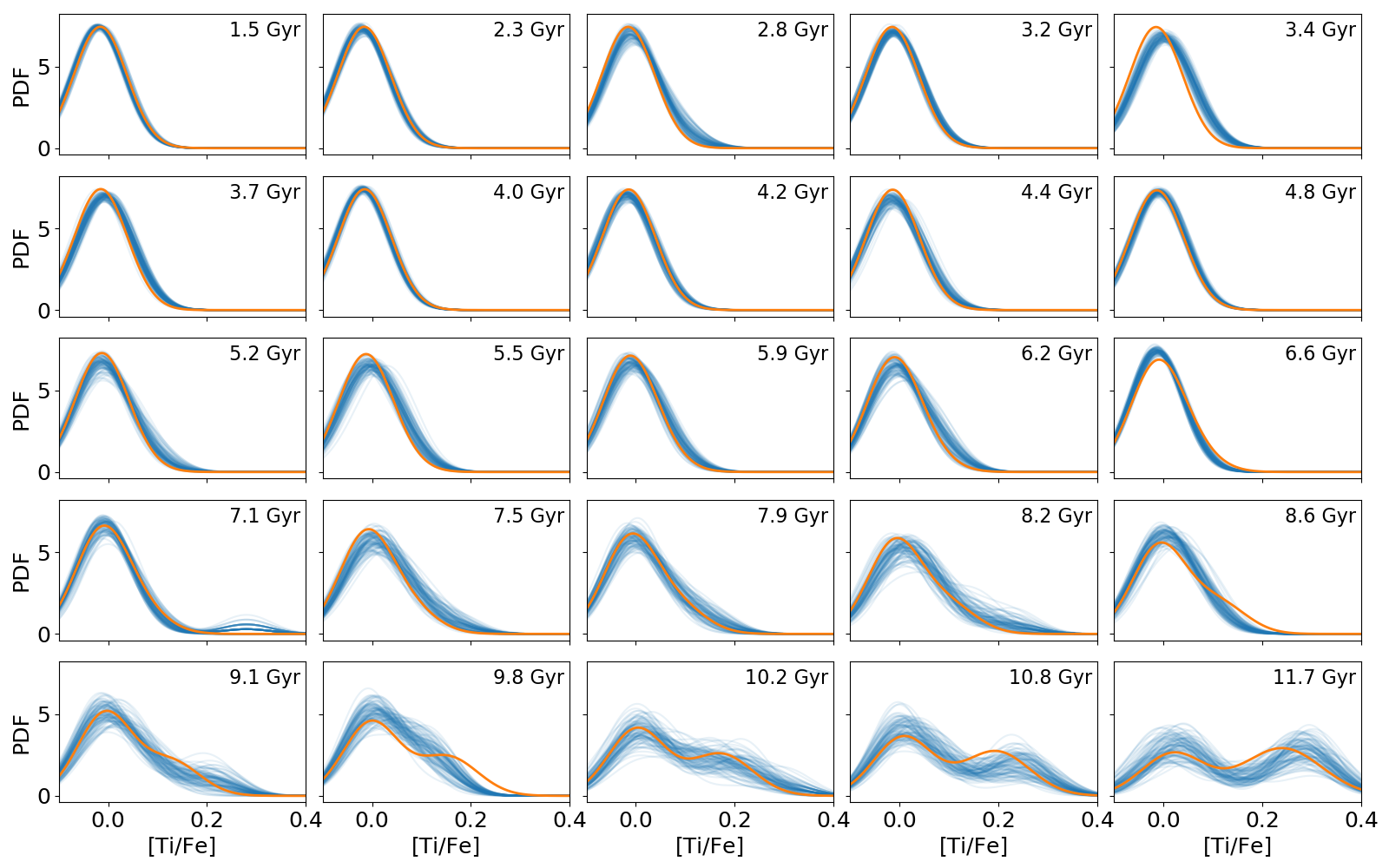}
\caption{Comparison of the \tife{} distribution of the B16 sample within each age bin to our Gaussian mixture model. Average age of the bin is shown in the top right corner of each subplot. Semitransparent blue lines represent kernel density estimation (KDE) distributions from 100 bootstrap samplings of the B16 \tife{} distribution within each age bin. Orange lines represent our empirical likelihood function derived from Gaussian mixture models with parameters from our power law fits (Equation~\ref{eq:gmm_param_fits} with best fit parameters from Table~\ref{tbl:gmm_param_fits}). Note, they are not direct fits to the distributions in blue. To incorporate the uncertainty in our M dwarf \tife{} measurements, we've convolved our empirical likelihoods with a Gaussian kernel with a standard deviation of 0.05 dex. We also use a Gaussian kernel with a standard deviation of 0.05 dex in the KDEs.\label{fig:gmm_fits_vs_kdes}}
\end{figure*}

Figure~\ref{fig:gmm_fits_vs_kdes} shows our Gaussian mixture distributions with means, variances, and weights given by Equation~\ref{eq:gmm_param_fits} with best fit parameters from Table~\ref{tbl:gmm_param_fits}. In order to use these distributions as likelihood functions for our planet-hosting M dwarf sample, we incorporate the uncertainty in our M dwarf \tife{} measurements by convolving the Gaussian mixture distributions with a Gaussian kernel with a standard deviation of 0.05 dex\footnote{Thanks to the distributive property of convolution and the fact that all distributions are Gaussian, this is equivalent to simply adding 0.0025 dex of additional variance to each Gaussian component}. These distributions represent the empirical probability of measuring a given \tife{} for given stellar age, incorporating both the intrinsic scatter in the \tife{}-age relation and the uncertainty in our M dwarf \tife{} measurements. The uncertainty in the B16 \tife{} measurements is much smaller than the intrinsic scatter within an age bin and is inherently included in this scatter. For qualitative comparison we also show kernel density estimation distributions from 100 bootstrap resamplings of the B16 \tife{} distribution within each age bin. We used a Gaussian kernel with a standard deviation of 0.05 dex. Evaluating the Gaussian mixture model at a measured \tife{} gives the likelihood as a function of age.

\section{Results}\label{sec:results}

\begin{figure}
\centering
\includegraphics[width=\linewidth]{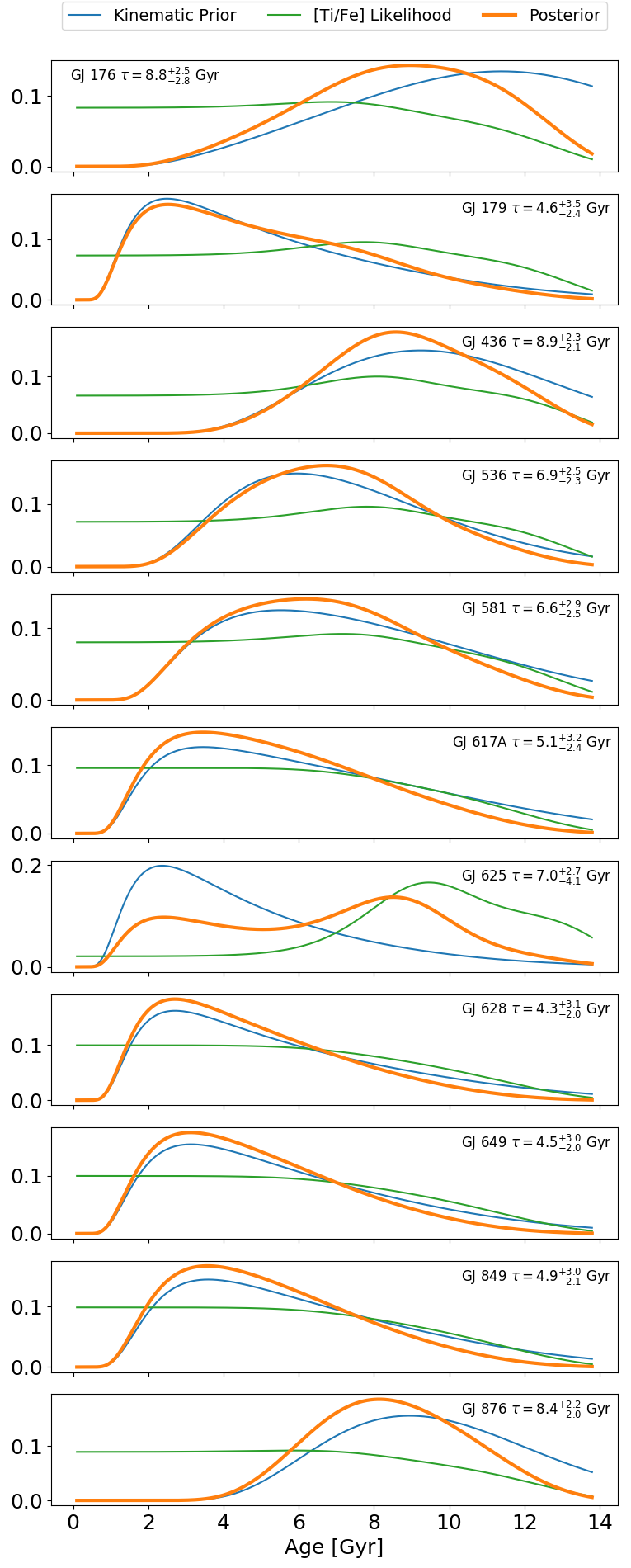}
\caption{Prior probability distribution, likelihoods, and posterior probability distribution for the age of each planet-hosting M dwarf in the CARMENES sample. Median and $\pm$1 sigma age estimates are listed in Table~\ref{tbl:ages}. \label{fig:planet_host_age_posteriors}}
\end{figure}

Figure~\ref{fig:planet_host_age_posteriors} shows the prior probability distribution, likelihoods, and posterior probability distribution for the age of each planet-hosting M dwarf in the CARMENES sample. Table~\ref{tbl:ages} lists the median of the posterior and $\pm$1 sigma uncertainties corresponding to 16th and 84th percentiles of the posterior. 

This approach preserves the astrophysical scatter in the relation between \tife{} and age, whereas fitting a parametric model directly to the \tife{}-age relation would assume all scatter is due to measurement uncertainty and would underestimate the uncertainty in predicted ages. However, this also means our age uncertainties may be overestimated as all scatter is taken to be astrophysical even though there is certainly measurement error (and likely systematic error) in both the \tife{} and age estimates. Nevertheless, we take the conservative approach of assuming all scatter is astrophysical and carry it through to our final age posteriors.

\section{Discussion}\label{sec:discussion}

\begin{figure}
\centering
\includegraphics[width=\linewidth]{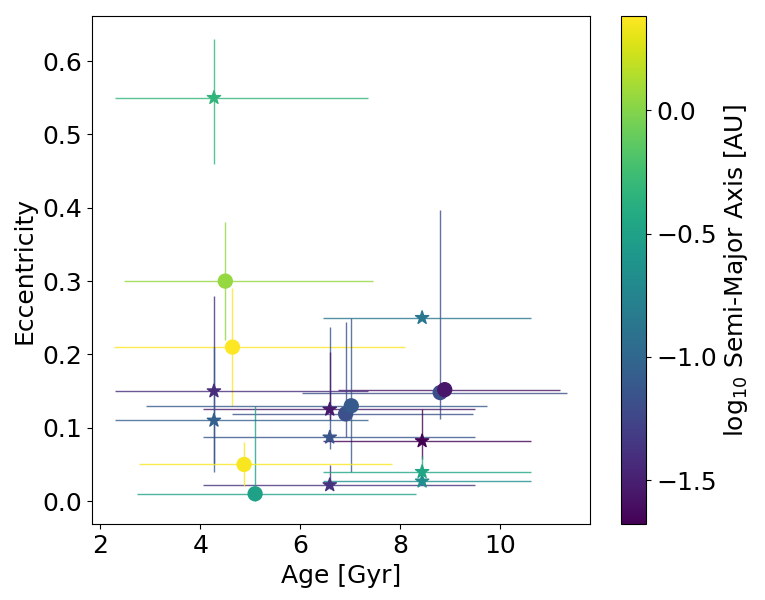}
\caption{Planet eccentricity as a function of stellar age, colored by semi-major axis. Planets in multi-planet systems are denoted as stars. \label{fig:age_vs_eccentricity}}
\end{figure}

Taken as an ensemble, we find that exoplanets orbit M dwarfs with a range of ages typical for the solar neighborhood. Our median age estimates range from 4~to~9~Gyr. Figure~\ref{fig:age_vs_eccentricity} shows the eccentricities of all planets in our sample compared with the age of the host star. The sample is too small to draw any significant conclusions about the eccentricity distribution as a function of age. However, we do find that there are a number of single planets on short-period orbits with low, but non-zero, eccentricities ($e\sim0.1\text{--}0.3$) at all ages. We also find that, on average, the sample is slightly Ti-enhanced. The mean \tife{} is $0.033 \pm 0.015$ dex and most M dwarfs in the sample have $\mtife{} > 0$. However, most are within their measurement error (0.05~dex) of the solar value and only one, GJ~625, has the distinct chemical signature of the metal-poor, alpha-rich thick disk.

\subsection{Tidal damping and migration}

In the absence of interactions with a third body, tides raised on both the planet and star are expected to damp out eccentricities and reduce the semi-major axis for planets orbiting within $\sim$0.2~AU of their host star \citep{Goldreich1966}. Tidal circularization likely explains the observed lack of highly eccentric ($e>0.5$) planets on close-in orbits noted in numerous studies \citep[e.g.,][]{Rasio1996,Kane2012,Kipping2013}. One way to determine whether or not a planet is currently undergoing tidal damping and migration is to estimate its tidal circularization timescale, $\tau_\mathrm{circ}$ \citep[Eq. 4 of][]{Jackson2008}. However, as \citet{Jackson2008} points out, tidal effects fall off rapidly with increasing semi-major axis. Therefore, it is important to model the coupled evolution of both eccentricity and semi-major axis to calculate the true time it takes to circularize an orbit.

\subsubsection{Simplified tidal circularization timescales}

For illustrative purposes, we used Eq.~4 in \citet{Jackson2008} (including the corrected numerical coefficient from \citealt{Jackson2009}) to calculate the simplified tidal circularization timescales for the planets in our sample. Calculating the tidal circularization timescale requires assuming a value for the tidal dissipation parameter $Q$, a unitless quality factor that is inversely proportional to tidal dissipation---smaller $Q$ results in stronger dissipation. We assumed a tidal dissipation parameter for the star of \Qs{}~=~$10^{5.5}$ and for the planet of \Qp{}~=~$10^{6.5}$ as suggested by \citet{Jackson2008}, although we note the tidal \Qp{} is no doubt different for each planet and likely depends on each planet's mass and structure. In our calculations, we used the planetary masses, semi-major axes, and eccentricities in Table~\ref{tbl:planets}. For planetary radii we used the empirical mass-radius-incident flux relation of \citet[][Eqs.~8~\&~9]{Weiss2013}. In one case, GJ~436~b, we had a measurement of the planetary radius from transit observations \citep{Maciejewski2014}. We estimated stellar masses and radii from the empirical absolute K-band magnitude relations of \citet[][Eq.~11 with coefficients from Tbl.~13]{Benedict2016} and \citet[][Eq.~5 with coefficients from Tbl.~1]{Mann2015}\footnote{We used the values from Table 1 of the erratum \citep{Mann2015erratum}.}, respectively. For both relations we used K-band magnitudes from 2MASS \citep{Skrutskie2006} and parallaxes from Gaia DR1 \citep{Gaia2016a,Gaia2016b,Lindegren2016}. We accounted for the Gaia zero-point offset, however, it has little effect as our stars are all nearby with parallaxes of order 100 mas.

\begin{figure}
\centering
\includegraphics[width=\linewidth]{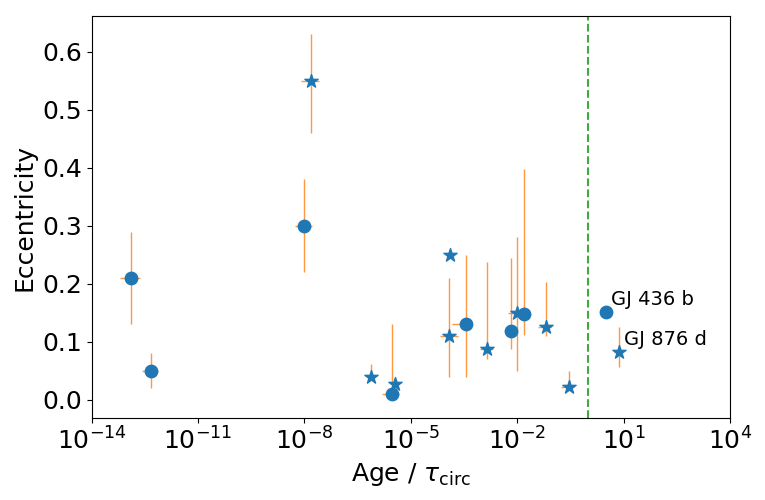}
\caption{Age divided by the simplified tidal circulation timescale, assuming \Qp{}=$10^{6.5}$, versus eccentricity. A green line indicates where age~=~$\tau_\mathrm{circ}$. Most planets in our sample have very long tidal circulation timescales such that tides are not expected to play a large role in the evolution of the planet even over the lifetime of the universe. Only two planets (labeled) have ages longer than their circularization timescales. \label{fig:tcirc_vs_eccentricity}}
\end{figure}

Figure~\ref{fig:tcirc_vs_eccentricity} shows the eccentricities of the planets in our sample versus the host star age divided by the tidal circularization timescale. Most planets in our sample have very long timescales, such that it is unlikely they underwent recent tidal evolution. Two planets, GJ~436~b and GJ~876~d, have tidal circularization timescales shorter than the age of their host star. It has been suggested that GJ~436~b has a massive, unseen companion that maintains its moderate eccentricity via Kozai interactions \citep{Beust2012}, so its short tidal circularization timescale may not be so surprising. GJ~876~d has three outer companions which are in a Laplace resonance \citep{Rivera2010}. GJ~876~d, the innermost planet in the system, is not expected to interact with the outer three planets \citep{Trifonov2018}, so its non-zero eccentricity is surprising given its short circularization timescale. We discuss GJ~436~b and GJ~876~d further in Sections~\ref{sec:GJ436}~\&~\ref{sec:GJ876}, respectively.

\subsubsection{Minimum \Qp{}}

\begin{figure}
\centering
\includegraphics[width=\linewidth]{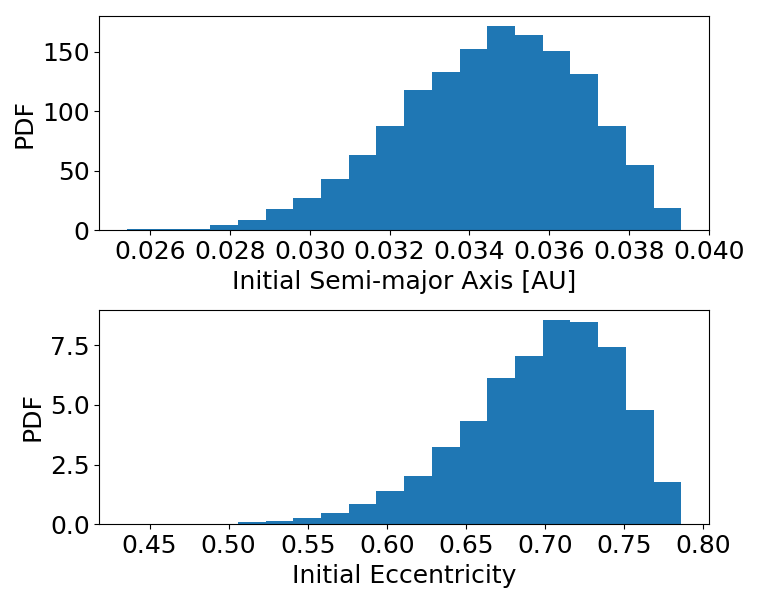}
\caption{Probability distributions for the initial eccentricity and semi-major axis of GJ~876~d based on integrating back the tidal evolution equations of \citet{Jackson2009} with current ages drawn from our age posterior. Here we assume $Q_\star = 10^{5.5}$ and $Q_p = 10^{6.5}$. \label{fig:GJ876d_e0_a0_Qs5.5_Qp6.5}}
\includegraphics[width=\linewidth]{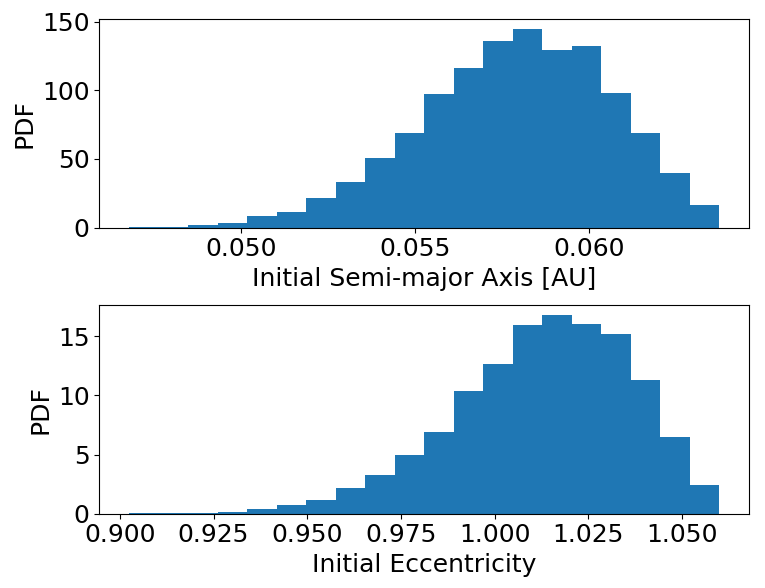}
\caption{Same as Figure~\ref{fig:GJ876d_e0_a0_Qs5.5_Qp6.5} but with $Q_\star = 10^{5.5}$ and $Q_p = 10^{5.5}$. \label{fig:GJ876d_e0_a0_Qs5.5_Qp5.5}}
\end{figure}

As pointed out by \citet{Jackson2008}, the simplified tidal circularization timescale ignores the coupled evolution of eccentricity and semi-major axis and can underestimate the true time to circularize. An alternative approach is to numerically integrate back the tidal evolution equations for both eccentricity and semi-major axis \citep[Eqs.~1~\&~2 of][]{Jackson2009} from the current age of the star to its formation. Doing so results in the initial eccentricity and semi-major axis of the planet's orbit, just after the protoplanetary disk dissipated, assuming no interactions with other bodies in the system. Since we have a posterior for the age of the star, we can determine the probability distribution for the initial eccentricity and semi-major axis of a planet if we assume a \Qp{} and \Qs{}. To do so, we integrate back the tidal evolution equations 10,000 times, each time drawing a random age from our age posteriors. We show an example of this for GJ~876~d in Figures~\ref{fig:GJ876d_e0_a0_Qs5.5_Qp6.5}~\&~\ref{fig:GJ876d_e0_a0_Qs5.5_Qp5.5} where we assume \Qs{}~=~$10^{5.5}$ and \Qp{}~=~$10^{6.5}$~and~$10^{5.5}$, respectively. Assuming a \Qp{} of $10^{6.5}$ results in a median initial eccentricity and semi-major axis of $e_i = 0.7$ and $a_i = 0.035$~AU. However, if we assume a \Qp{} of only $10^{5.5}$, the median initial eccentricity exceeds 1. Therefore, we can constrain the \Qp{} of GJ~876~d to be greater than ${\sim}10^{5.5}$.

\begin{figure}
\centering
\includegraphics[width=\linewidth]{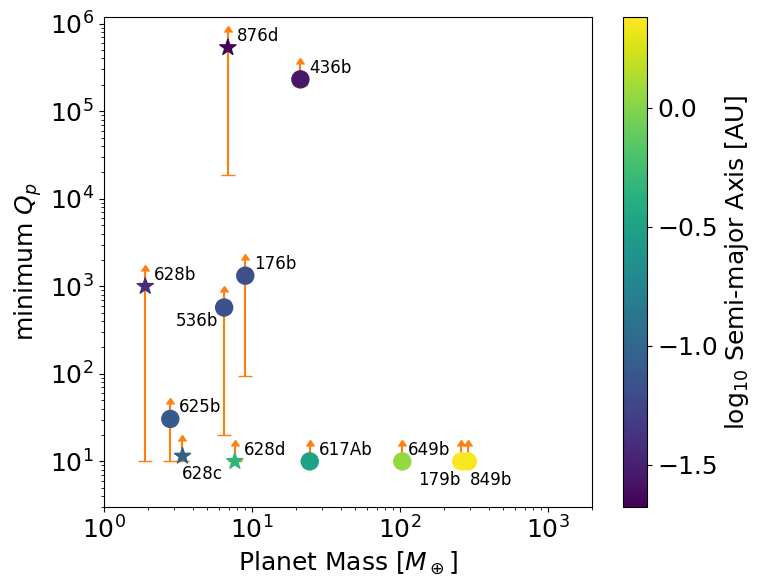}
\caption{Minimum $Q_p$ versus planet mass, colored by orbital period. Stars indicate planets in multi-planet systems. Lower error bars are calculated by assuming the planet radius and eccentricity are overestimated by 1$\sigma$. Planets known to be gravitationally interacting with other planets in the system have been excluded (GJ~581~b,c,e and GJ~876~b,c,e). \label{fig:minQps}}
\end{figure}

Following this approach, we can use our host star ages to estimate the minimum \Qp{} possible for each planet by stepping through possible \Qp{} values until the initial eccentricity exceeds one. For each planet, we estimate the initial eccentricity probability distribution for 100 \Qp{} values space logarithmically from 10 to $10^7$. We take the maximum \Qp{} at which the median initial eccentricity is greater than one as the minimum possible \Qp{} for the planet. We hold the stellar tidal dissipation parameter fixed at $10^{5.5}$. For planets that are susceptible to tidal effects around M dwarfs (i.e. on close-in orbits), the effect of the stellar tide is negligible. The minimum \Qp{} values are shown in Figure~\ref{fig:minQps}. Again, this assumes only tidal interactions with the host star so the results are not necessarily meaningful for dynamically interacting multi-planet systems. We've excluded from the figure, planets known to be gravitationally interacting with a companion (GJ~581~b,c,e and GJ~876~b,c,e). To assess the sensitivity of our minimum \Qp{} estimates to measurement uncertainties, we redid the analysis using planet radii and eccentricities that were smaller by 1$\sigma$. For radii, we used the RMSE of the mass-radius relation \citep[1.41~$R_\earth$ for $M<150M_\earth$, 1.15~$R_\earth$ for $M \ge 150M_\earth$,][]{Weiss2013}, except for GJ~436~b where we used the radius uncertainty quoted by \citet{Maciejewski2014}. For eccentricities, we used the lower uncertainties listed in Table~\ref{tbl:planets}.

The tidal dissipation parameter \Qp{} is poorly constrained even for the planets in the Solar System. Gas giants like Jupiter and Saturn have \Qp{} values somewhere around $10^5$--$10^6$ \citep{Goldreich1966,Ioannou1993,Ogilvie2004}. Although, \citet{Lainey2009,Lainey2012,Lainey2017} suggest Jupiter and Saturn's tidal \Qp{} are much lower, around 35000 and 2500, respectively. Neptune and Uranus both have \Qp{} values around $10^4$ \citep{Goldreich1966,Zhang2008}. Rocky planets have low \Qp{} values around 10-200 \citep{Goldreich1966,Murray2000,Lainey2007,Henning2009}. Because of the large separation between gas giant and terrestrial \Qp{} values, they can be used to differentiate between rocky and gaseous planets \citep{Barnes2015}. Since we can only provide a lower bound on \Qp{}, we cannot place strict constraints on the composition of these planets, though a high minimum \Qp{} might suggest the planet is more like a gas-giant than a rocky planet. It is also important to note that tidal evolution is strongly dependent on the radius of the planet ($1/\tau_\mathrm{circ} \propto R_\mathrm{p}^5$). For all but one planet we assume a radius based on mass and incident flux. Therefore, a high minimum \Qp{} does not necessarily rule out a rocky composition. Rather, a high minimum \Qp{} indicates that a planet could be affected by tidal interactions, if its true \Qp{} is not well above our lower limit and its true radius is not much less than that inferred from empirical mass-radius relations, which are know to have significant scatter \citep{Wolfgang2015,Wolfgang2016}. A small minimum \Qp{} simply means the planet is insensitive to tidal effects on timescales of the stellar lifetime, e.g., if it obits at a large semi-major axis. 

It is also important to note that the tidal evolution equations are only valid under the assumption that the planet's orbital period is shorter than the star's rotational period. While this may not be valid for all systems in this study, it is valid for the most interesting systems, those with short-period planets ($P < 10$ days) around old stars. Field early-to-mid-M dwarfs typically have rotation periods $>$ 10 days \citep{McQuillan2013} and kinematically old mid-M dwarfs typically rotate with periods $>$ 70 days \citep{Irwin2011,Newton2016}.

\subsection{Individual systems}

In the following we discuss individual systems in more detail.

\subsubsection{GJ 176}
GJ~176 hosts a super-Earth in an 8.8 day orbit originally discovered by \citet{Forveille2009}. \citet{Trifonov2018} published updated parameters for GJ~176~b, incorporating 23 new CARMENES observations and confirming a $M \sin i \approx 9 M_\earth$ planet in a $\sim$8.8 day orbit. They report a mildly eccentric orbit with $e = 0.148^{+0.249}_{-0.036}$.

\citet{Eggen1998} proposed that GJ~176 is a member of the moving group HR 1614 based on its kinematics. \citet{Feltzing2000} estimated HR 1614 to be about 2~Gyr old. However, our analysis suggests GJ~176 is an older star with an age of $8.8^{+2.5}_{-2.8}$~Gyr and rules out ages less than 2~Gyr at the $3 \sigma$ level. Furthermore, \citet{DeSilva2007} spectroscopically analyzed 18 proposed members of HR 1614 and found the cluster to be metal rich with $\log\varepsilon_\mathrm{Fe} = 7.77 \pm 0.033$~dex. \citet{DeSilva2007} also found that 4 out of the 18 stars they studied had lower metallicities with $\log\varepsilon_\mathrm{Fe}$~=~7.44--7.55~dex and deviated from
the cluster mean abundances in all elements except the $n$-capture elements, suggesting they are not members of HR 1614. We measure an iron abundance for GJ~176 of only $\log\varepsilon_\mathrm{Fe} = 7.5 \pm 0.1$~dex, further suggesting it is not a member of HR 1614.

GJ~176~b has the second highest minimum \Qp{} (${\sim}10^3$) of all single-planet systems in our sample. At a minimum mass of ${\sim}9\,M_\earth$ GJ~176~b is likely more similar to Uranus and Neptune than to a massive, rocky super-Earth. This is supported by our minimum \Qp{} for GJ~176~b which is close to the \Qp{} of Uranus and Neptune. Orbiting at only 0.066~AU, GJ~176~b has likely undergone some tidal evolution. Assuming a radius of 3.5 $R_\earth$ and a \Qp{} of $10^4$, we estimate GJ~176~b started with a high initial eccentricity of $e_i \approx 0.7$ and migrated in from an initial semi-major axis of $a_i \approx 0.1$~AU.

\subsubsection{GJ 179}
GJ~179 is one of the few M dwarfs known to host a Jupiter-analog. GJ~179~b is a Jupiter-mass planet ($M \sin i = 0.82 M_{Jup}$) in a slightly eccentric ($e=0.21\pm0.08$) 6.3 year orbit \cite{Howard2010}. Despite a slight Ti-enhancement, the kinematics of GJ~179 suggest a moderate age of $4.6^{+3.5}_{-2.4}$~Gyr, but with a long tail of probability up to older ages.

Orbiting at such a large distance from its host star, GJ~179~b is not expected to be affected by tidal interactions with the host star.

\subsubsection{GJ 436}\label{sec:GJ436}

GJ~436 was the second M dwarf found to host an exoplanet \citep{Butler2004}. GJ~436~b has roughly the same radius, mass, and density of Neptune, but orbits with a period of only 2.6 days \citep{Maciejewski2014}. \citet{Butler2004} originally estimated that GJ~436 is more than 3 Gyr old based in its kinematics and chromospheric activity. By combining kinematics with Ti-enhancement, we constrain the age to be $8.9^{+2.3}_{-2.1}$ Gyr, making GJ~436 the oldest planet host in our sample (with the potential exception of GJ~625, see Section~\ref{sec:GJ625}).

The old age of GJ~436 is surprising considering that it hosts a short-period planet in an eccentric orbit \citep[$e=0.152\pm0.009$,][]{Trifonov2018}. \citet{Bourrier2018} recently reported that the orbit of GJ~436~b is not only eccentric, but also nearly perpendicular with the spin axis of the star.

One scenario that has been proposed to explain the eccentricity of GJ~346~b is interaction with a third body \citep{Maness2007,Demory2007,Mardling2008,Ribas2008}. \citet{Tong2009} investigated the possible locations of a dynamical companion in either a resonant and non-resonant orbit and argued that the eccentricity of GJ~436~b can not be maintained by either a nearby or distant companion. \citet{Batygin2009} confirmed that, for most scenarios, the presence of a second planet does not keep GJ~436~b from rapidly circularizing. However, they found that under certain initial conditions where the eccentricities of the two planets are locked at a quasi-stationary point, the eccentricity damping of GJ~436~b can be extended to $\sim$8~Gyr.

\citet{Beust2012} put forward another hypothesis, suggesting that GJ~436~b originally orbited  at a larger semi-major axis and migrated to its current orbit via Kozai migration induced by a distant perturber. In this scenario, the eccentricity damping of GJ~436~b can be delayed by several Gyr. \citet{Bourrier2018} estimated the age of GJ~436 to be $\sim$5~Gyr based on its rotation period of 44~days and found that Kozai migration could explain both the eccentricity and obliquity of GJ~436~b.

\citet{Morley2017} found that additional interior heat from tidal dissipation is required to explain the observed thermal emission of GJ~436~b. They were able to constrain the tidal \Qp{} of GJ~436~b to $2 \times 10^5$--$10^6$. This agrees well with our minimum \Qp{} estimate of $\sim$10$^5$. With a \Qp{}~$> 10^5$, the tidal dissipation is weak enough that an unseen third body is not required to explain the non-zero eccentricity. However, it does mean that the orbit of GJ~436~b has been significantly altered by tidal effects over its lifetime. Assuming a \Qp{}~=~$10^6$, GJ~436~b would have initially orbited with an eccentricity around $\sim$0.8 at a distance of $\sim$0.05~AU. However, this scenario does not explain the high obliquity of the current orbit reported by \citet{Bourrier2018}.

\subsubsection{GJ 536}

GJ~536 hosts a super-Earth planet in a 8.7 day orbit \citep{SuarezMascareno2017a}. Using additional RV data from the CARMENES survey, \citet{Trifonov2018} refined GJ 536 b's mass estimate to $M \sin i = 6.52^{+0.69}_{-0.40} M_\earth$ and the eccentricity of the orbit to $e = 0.119^{+0.125}_{-0.032}$. We estimate the age of GJ~536 to be $6.9^{+2.5}_{-2.3}$ Gyr.

GJ~536~b is very similar to GJ~176~b. Both orbit at $\sim$0.066~AU with similar eccentricities, 0.15 and 0.12. GJ~176~b has a slightly higher minimum mass than GJ~536~b, 9~$M_\earth$ versus 6.5~$M_\earth$. We estimate the minimum \Qp{} of both planets to be ${\sim}10^3$. In the absence of interactions with other unseen planets in the system, GJ~176~b and GJ~536~b are likely similar mini-Neptune planets with extensive gaseous atmospheres. Both are also likely to have undergone tidal circularization and migration. Assuming a \Qp{} of $10^4$, we estimate GJ~536~b initially orbited at $\sim$0.08~AU with an eccentricity of $\sim$0.5.

\subsubsection{GJ 581}

GJ~581 hosts three bona fide planets: GJ~581~b \citep{Bonfils2005b}, GJ~581~c \citep{Udry2007}, and GJ~581~e \citep{Mayor2009}. The three planets orbit in a very compact configuration with semi-major axes between 0.029~and~0.074~AU. \citet{Trifonov2018} used an $N$-body model to show that all three planets are dynamically interacting and are in a stable configuration where each planets semi-major axis is constant, but their eccentricities oscillate on timescales of 50 and 500 years. Since all three planets are interacting, our minimum \Qp{} estimates are invalid. \citet{Trifonov2018} showed that this configuration is stable for at least 10~Myr. Our age estimate for GJ~581 of $6.6^{+2.9}_{-2.5}$ Gyr suggests that these compact, interacting systems can be stable for several Gyr. This is further supported by the apparent ubiquity of these ``compact multiples'' \citep{Muirhead2015}.

\subsubsection{GJ 617 A}

HD~147379 (GJ~617~A) was the first star discovered to host a planet by the CARMENES survey \citep{Reiners2018}. GJ~617A~b has a minimum mass of $M_p \sin i \sim 25 M_\earth$ and orbits in a nearly circular $\sim$0.3 AU orbit. As such, GJ~617A~b is unlikely to be strongly affected by tidal interactions with its host star.

\citet{Vican2012} estimated the age of GJ~617~A to be $\sim$1~Gyr based on chromospheric activity and X-ray flux. However, they used the activity-age relations of \citet{Mamajek2008} which were only calibrated down to early K dwarfs ($B-V < 0.9$~mag) and GJ~617~A is a late-K/early-M with $B-V = 1.34$. We estimate a slightly older chemo-kinematic age for GJ~617~A of $5.1^{+3.2}_{-2.4}$~Gyr.

\citet{Reiners2018} found an additional peak in the periodogram of GJ~617~A corresponding to a period of 21 days which they attribute to the roation period of the star. \citet{Pepper2018} confirmed a 22 day rotation period based on 3304 KELT observations \citep{Pepper2007}. \citet{Agueros2018} recently measured rotation periods for 12 K and M dwarfs in the 1.34~Gyr old cluster NGC~752. They found that late-K dwarfs similar in mass to GJ~617~A rotate with a rotation period of $\sim$15~days. Assuming a simple Skumanich-like evolution ($\tau \propto p_\mathrm{rot}^2$), a rotation period of 22~days for GJ~617~A would suggest an age of $\sim$3~Gyr, in rough agreement with our chemo-kinematic estimate.

\subsubsection{GJ 625}\label{sec:GJ625}
GJ~625 was only recently discovered to host a super-Earth orbiting at the inner edge of the habitable zone \citep{SuarezMascareno2017b}. A rocky planet orbiting at such a close distance to its host star is expected to circularized on very short timescales. Indeed, the eccentricity of GJ~625~b is consistent with zero; $e = 0.13^{+0.12}_{-0.09}$. GJ~625~b is not likely to be currently undergoing tidal migration, although it may have in the past.

GJ~625 is a peculiar case where its kinematics strongly favor a young age, however, its low \feh{} and high \tife{} abundances are similar to older thick disk members. This leads to a combined age posterior that is bimodal and has significant probability at essentially all ages. The median and $\pm 1\sigma$ values of the posterior are $7.0^{+2.7}_{-4.1}$~Gyr. Estimating the age from the kinematic prior alone yields $3.9^{+3.3}_{-1.9}$~Gyr. Whereas ignoring the kinematic prior and assuming a flat prior results in an age estimated from \tife{} alone of $9.5^{+2.5}_{-3.0}$~Gyr.

We can turn to other indications of an M dwarf's age to argue in favor of either the young or old interpretation of the age of GJ~625. A common indicator for the rough age of an M dwarf is its rotation period. \citet{SuarezMascareno2017b} estimate the rotation period of GJ~625 to be $P = 77.8 \pm 5.5$~days. The relation between rotation period and age for M dwarfs is not well understood. However, studies of young open clusters and field M dwarfs suggest M dwarfs, like solar-type stars, spin down over time as magnetized stellar winds carry away angular momentum \citep{Irwin2007,Irwin2011,McQuillan2013,Newton2016,Rebull2017,Douglas2017}. \citet{Newton2016} found that field mid-M dwarfs like GJ~625 show a bimodal distribution in rotation period with peaks at $\sim$1 and $\sim$100 days. The evolutionary link between these two populations is not clear, however, \citet{Newton2016} found that M dwarfs with rotation periods $>$70 days are kinematically consistent with an old population with an average age of 5~Gyr. GJ~625's slow $\sim$80 day rotation is similar to that of the slowest rotating (and presumably oldest) stars in the field of similar mass \citep{Newton2017}, and is therefore more consistent with the older peak of our age posterior.

\citet{Montes2001} lists GJ~625 as a possible member of the young Ursa Major moving group which would imply and age of only $\sim$0.5 Gyr (although \citealt{Montes2001} do list it with the caveat that it fails both their peculiar velocity and radial velocity criteria). We can rule out membership based on GJ~625's long rotation period, low activity \citep[$\log_{10}(R'_{HK}) = -5.5\pm0.2$,][]{SuarezMascareno2017b}, and chemical dissimilarity \citep{Tabernero2017}. We also note that the BANYAN~$\Sigma$ web tool\footnote{\url{http://www.exoplanetes.umontreal.ca/banyan/banyansigma.php}} lists a 0\% probability that GJ~625 is a member of the Ursa Major moving group and 99.9\% probability it is a field star \citep{Gagne2018}.

\subsubsection{GJ 628}
\citet{Wright2016} used archival HARPS spectra to discover three potentially rocky planets around GJ~628 (Wolf~1061), with one planet, GJ~628~c, orbiting within the habitable zone. \citet{Astudillo-Defru2017} rule out a third planet orbiting at 67 days as originally proposed by \citet{Wright2016}, but find significant evidence for a third planet orbiting at 217 days.

GJ~628 is one of a couple cases where a solar \tife{} estimate results in very broad likelihood and the posterior is dominated by the kinematic prior which favors younger ages. Based on this, we estimate an age for GJ~628 of $4.3^{+3.1}_{-2.0}$~Gyr. However, \citet{Astudillo-Defru2017} claim a rotation period of 95~days for GJ~628, which is supported by the photometric monitoring presented in \citet{Kane2017}. Such a long rotation period suggests an age $>$~5~Gyr, as discussed in the previous section. If a strict gyrochronological relation does exists for M dwarfs, this long rotation period is inconsistent with our age estimate.

We estimate a minimum \Qp{}~$\sim 10^3$ for the inner most planet, GJ~628~b. Such a high minimum \Qp{} would suggest the $M_p \sin i \sim 2 M_\earth$ planet is not rocky, but in fact has a large gaseous envelope. However, GJ~628~b and GJ~628~c have eccentricities that are consistent with zero within measurement error. If we assume their radii and eccentricities are overestimated by 1$\sigma$, we can no longer constrain the minimum \Qp{} to be greater than 10. Therefore, given that they are both likely rocky and orbit within 0.1~AU of their host star, they likely were tidally circularized not long after their primordial disk dissipated.

\citet{Montes2001} lists GJ~628 as a member of the young \citep[125~Myr,][]{Stauffer1998} Pleiades moving group. However, the BANYAN~$\Sigma$ web tool lists a 0\% probability that GJ~628 is a member of the Pleiades moving group and 99.9\% probability it is a field star.

\subsubsection{GJ 649}
GJ~649 hosts one known planet with a minimum mass similar to Saturn, $M_p \sin i \sim 100 M_\earth$ \citep{Johnson2010b}. With a semi-major axis of 1.1~AU, the orbit of GJ~649~b is not expected to be influenced by tidal effects.

GJ~649 is another case where a near solar \tife{} does little to constrain the age of the star---other than ruling out the oldest ages---and the kinematics favor younger ages. We estimate an age of $4.5^{+3.0}_{-2.0}$~Gyr for GJ~649.

\subsubsection{GJ 849}
GJ~849 hosts a roughly Jupiter mass planet in a nearly circular, $\sim$5-year orbit \citep{Butler2006}. Orbiting at over 2~AU, GJ~849~b is not expected undergo any tidal circularization or migration. \citet{Butler2006} noted a linear trend in their RV time series data, suggesting a possible second planet in the system on an even longer orbit. With RV measurements spanning 17 years, \citet{Feng2015} find strong evidence for a second planet with $M \sin i \approx M_J$ orbiting with a period of $15.1 \pm 1.1$~years.

Like GJ~628 and GJ~649, GJ~849 has nearly solar \tife{} and kinematics that skew the age posterior to younger ages. Based on this, we estimate the age of GJ~849 to be $4.9^{+3.0}_{-2.1}$~Gyr.

\subsubsection{GJ 876}\label{sec:GJ876}

The planetary system around GJ~876 is a benchmark system for studying the formation and migration of planets in compact systems. GJ~876 hosts a total of four know planets, three of which are in a Laplace 1:2:4 mean-motion resonance \citep{Rivera2010}. The resonance is chaotic, but expected to be stable on timescales of at least 1 Gyr \citep{Rivera2010,Marti2013}. The old age we infer for GJ~876 ($8.4^{+2.2}_{-2.0}$~Gyr) suggests such chaotic resonances can be stable for several Gyr, assuming the planets migrated into this configuration soon after their formation \citep{Batygin2015}.

Even though the innermost planet, GJ~876~d, is not in the resonant chain with the other three planets, our old age for GJ~876 has some interesting implications for its past migration. Originally, \citet{Rivera2010} estimated planet d orbited with an unusually high eccentricity ($e=0.207\pm0.055$) for a planet that orbits at only 0.02~AU with a period of about 2~days. Recently, \citet{Trifonov2018} and \citet{Millholland2018} reanalyzed the system by fitting dynamical $N$-body models to existing and new RV data. Both analyses revised the eccentricity of planet d down to values more consistent with a circular orbit. \citet{Trifonov2018} report a best fit eccentricity of $e=0.082^{+0.043}_{-0.025}$ while \citet{Millholland2018} report a best fit eccentricity of $e=0.057\pm0.039$. Using the \citet{Trifonov2018} estimate for the eccentricity, we constrain the minimum \Qp{} of GJ~876~d to be $> 5 \times 10^5$. Such a high \Qp{} is surprising for a $\sim 7 M_\earth$ planet. Our minimum \Qp{} for GJ~876~d is an order of magnitude higher than the \Qp{} of Uranus and Neptune and three orders of magnitude greater than the \Qp{} of terrestrial bodies.

A few scenarios could explain the non-zero eccentricity of GJ~876~d. One explanation is that the planet has a peculiar structure that is very inefficient at dissipating tidal energy. Another explanation is that GJ~876~d originally orbited with a larger semi-major axis where it gravitationally interacted with the outer three planets and, at some point in its history, fell out of a chaotic resonance with the outer planets and migrated in. A third scenario is that there is another, unseen planet in the system that is gravitationally interacting with GJ~876~d.

We note that the eccentricity of GJ~876~d has undergone downward revision recently. If additional observations in future studies lead to the further downward revision and the conclusion that the orbit is essentially circular, it may not be necessary to invoke one of the above scenarios to explain the orbit of GJ~876~d. In that case, our minimum \Qp{} will be overestimated. However, given its proximity its host star and our age estimate of $8.4^{+2.2}_{-2.0}$~Gyr for the system, GJ~876~d likely has undergone significant tidal circularization and migration in its lifetime.

\citet{Montes2001} list GJ~876 as a member of the young Pleiades moving group which is inconsistent with our old age estimate for the star. The BANYAN~$\Sigma$ web tool gives a 0\% probability that GJ~876 is a member of the Pleiades moving group and a 81.7\% probability it is a field star. Interestingly, it also gives a 18.3\% probability that GJ~876 is a member of the Beta Pictoris moving group.

\section{Summary}\label{sec:summary}

We used a sample of well-studied FGK stars to develop a data-driven approach to estimate the ages of field stars from their composition and kinematics within a Bayesian framework. Our method relies on astrophysical trends between stellar ages, UVW space velocities, and titanium enhancement \tife{}. We applied our method to 11 exoplanet-hosting M dwarfs, making use of recent advancements in the detailed chemical analysis of M dwarfs \citep{Veyette2016b,Veyette2017}. We list our exoplanet host ages in Table~\ref{tbl:ages}.

Tidal effects are expected to circularize the orbits of short-period planets around M dwarfs. However, we find a number of close-in planets ($a < 0.1$~AU) with mildly eccentric orbits ($e \sim 0.1$) in fact orbit relatively old stars with ages around 8~Gyr. For these stars, we can constrain the minimum tidal \Qp{} possible that can explain the current eccentricity, semi-major axis, and age of the system.

We find that GJ~176~b and GJ~536~b, two short-period mini-Neptune planets on similar orbits, have similar minimum \Qp{} values of $\sim$10$^3$, suggesting mini-Neptune planets have \Qp{} values closer to those of the ice giants than the terrestrial planets in our Solar System. We estimate the ages of the host stars of these systems to be $8.8^{+2.5}_{-2.8}$ and $6.9^{+2.5}_{-2.3}$ Gyr and find both planets likely have undergone tidal migration and circularization and initially orbited farther from their host star with eccentricities $>$ 0.5.

We estimate an age of $8.9^{+2.3}_{-2.1}$~Gyr and a minimum \Qp{} of $\sim$10$^5$ for GJ~436~b. Our \Qp{} limit agrees well with \citet{Morley2017} who used the observed thermal emission of GJ~436~b to constrain its tidal heating and \Qp{}. With such a high \Qp{}, a gravitationally interacting third body in the system is not required to explain the non-zero eccentricity of GJ~436~b, as suggested by numerous authors. However, this scenario does not explain the high obliquity of the orbit reported by \citet{Bourrier2018}.

We estimate an old age of $8.4^{+2.2}_{-2.0}$~Gyr for GJ~876, which hosts three outer planets in a Laplace resonance and a fourth inner planet that is not expected to interact with the resonance. This old age is surprising given that the innermost planet, GJ~876~d, orbits at only 0.02~AU and has a nonzero eccentricity. We estimate a very high minimum \Qp{} of $5\times10^5$ for the $\sim$7~$M_\earth$ planet which suggests either that (1) GJ~876~d has a peculiar structure that is very inefficient at dissipating tidal energy, (2) GJ~876~d originally orbited farther out where it interacted with the resonant chain and at some point fell out of resonance and migrated in, or (3) there is another unseen companion that is interacting with GJ~876~d and maintaining its nonzero eccentricity.

\acknowledgements
We thank the anonymous referee for their thoughtful comments and useful suggestions.

The authors would like to thank the CARMENES consortium for making their spectra publicly available. We also thank Paul Dalba, Julie Skinner, and Todd Henry for valuable discussions during the preparation of this manuscript.

This material is based upon work supported by the National Science Foundation under Grant No. 1716260.  This work was partially supported by a NASA Keck PI Data Award, administered by the NASA Exoplanet Science Institute.  This research made use of the NASA Exoplanet Archive, which is operated by the California Institute of Technology, under contract with the National Aeronautics and Space Administration under the Exoplanet Exploration Program.

This work made use of data from the European Space Agency (ESA)
mission {\it Gaia} (\url{https://www.cosmos.esa.int/gaia}), processed by
the {\it Gaia} Data Processing and Analysis Consortium (DPAC,
\url{https://www.cosmos.esa.int/web/gaia/dpac/consortium}). Funding
for the DPAC has been provided by national institutions, in particular
the institutions participating in the {\it Gaia} Multilateral Agreement.

This publication made use of data products from the Two Micron All Sky Survey, which is a joint project of the University of Massachusetts and the Infrared Processing and Analysis Center/California Institute of Technology, funded by the National Aeronautics and Space Administration and the National Science Foundation.

\software{\texttt{isochrones} \citep{Morton2015}, \texttt{emcee} \citep{Foreman-Mackey2013}, \texttt{numpy} \citep{numpy}, \texttt{scipy} \citep{scipy}, \texttt{matplotlib} \citep{matplotlib}}

\bibliographystyle{aasjournal}
\bibliography{bib}

\appendix
\section{Ages of the B16 stars}\label{fgkages}

The \texttt{isochrones} package takes a number of observables along with estimates of their Gaussian uncertainties and compares them to interpolated stellar evolution models to estimate stellar parameters. In an attempt to reduce the effects of systematic differences between the stellar evolution models and spectroscopically derived parameters, we opted to include as many observables as possible with realistic absolute errors. The observables we used are \teff{}, \mh{}, parallax, $B_\mathrm{T}$ magnitude, $V_\mathrm{T}$ magnitude, $J$ magnitude, $H$ magnitude, and $K_\mathrm{S}$ magnitude.

For \teff{}, we adopted the values from B16 but included a $-$39 K offset. The offset corresponds to the mean difference between the spectroscopic \teff{} from B16 and the \teff{} from optical interferometry \citep{Boyajian2013} for stars with both measurements. Spectroscopic temperatures are not necessarily equivalent to the effective temperatures used by stellar evolution models---defined as $\mteff{} = \left(L / 4 \pi R^2 \sigma \right)^{1/4}$, where $L$ is the luminosity and $R$ is the radius of star. We also found that when excluding the spectroscopic parameters and only fitting to the parallax distance and observed magnitudes, the best fit model \teff{} values were on average 40 K cooler than the spectroscopic temperatures. This combined with the comparison to interferometric \teff{} measurements suggest that the spectroscopic \teff{} are slightly overestimated. The quoted statistical uncertainty from B16 is $\pm$25 K. We, however, used a conservative estimate of $\pm$80 K uncertainty in \teff{} which corresponds to the RMS scatter between spectroscopic and interferometric \teff{} measurements.

For metallicity, we used the \mh{} values from B16. The MIST models assume scaled solar abundances, parameterized by a single metallicity parameter. The \mh{} values of B16 represent the best-fitting solar-scaled abundances, before tuning individual abundances, which is more akin to how metallicity is treated in the MIST models compared to assuming \feh{} as the metallicity. The B16 quoted statistical uncertainty on \mh{} is 0.01 dex. We adopt an uncertainty of 0.1 dex in order to account for various systematic errors such as differences in the assumed solar abundances of B16 \citep{Grevesse2007} and the MIST models \citep{Asplund2009} and systematic error from the simple assumption of scaled solar abundances.

We cross-matched the B16 sample with the HIPPARCOS \citep{hipparcos} and Gaia DR1 TGAS \citep{Gaia2016a,Gaia2016b,Lindegren2016} catalogs. When available, we used Gaia parallaxes and quoted uncertainties. If Gaia data was not available, we used HIPPARCOS parallaxes with quoted uncertainties.

We included five magnitudes with their quoted uncertainties. We included $B_\mathrm{T}$ and $V_\mathrm{T}$ magnitudes from the Tycho-2 catalog \citep{Hog2000} and $J$, $H$, and $K_\mathrm{S}$ magnitudes from 2MASS \citep{Skrutskie2006}.

We made two changes to \texttt{isochrones} package. First, we implemented a Jefferys prior for $A_V$ within the bounds 0.001 $<$ $A_V$ $<$ 1. Second, we implemented an \texttt{Isochrone} class for the MIST models as opposed to the default \texttt{FastIsochrone} class. We found that the interpolation scheme used in the \texttt{FastIsochrone} class produced strange artifacts such as striations in 2D marginalized posteriors. The \texttt{Isochrone} class uses the \texttt{scipy.interpolate.LinearNDInterpolator} function to interpolate the models. To speed up computing, we calculate the Delaunay triangulation only for age $>$ 0.1 Gyr, \mh{} $>$ $-$1, and 0.5 $<$ $M/M_\sun$ $<$ 1.5 which encompasses our entire FGK sample after making the cuts described in Section~\ref{sec:likelihood}.

\begin{figure*}
\centering
\includegraphics[width=\linewidth]{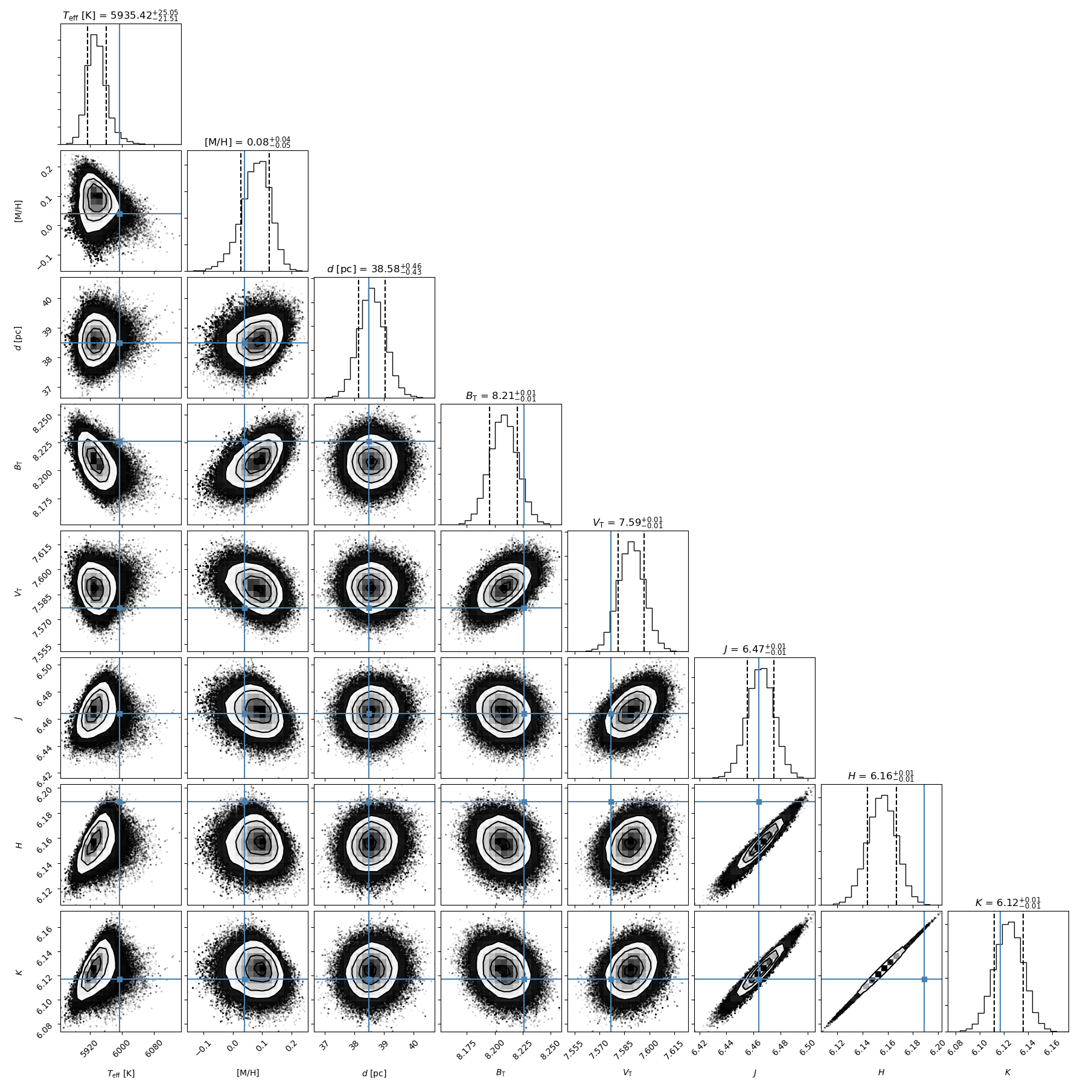}
\caption{Corner plot showing distributions of modeled observables from sampling the posterior for HD 105. Blue crosshairs indicate measured values. The uncertainties on the observables are 80 K in \teff{}, 0.1 dex in \mh{}, 0.015 mag in $B_\mathrm{T}$, 0.001 mag in $V_\mathrm{T}$, 0.02 mag in $J$, 0.023 mag in $H$, and 0.02 mag in $K_\mathrm{S}$. The observables are well reproduced by the model to within measurement uncertainties. \label{fig:isochrones_in}}
\end{figure*}

\begin{figure*}
\centering
\includegraphics[width=\linewidth]{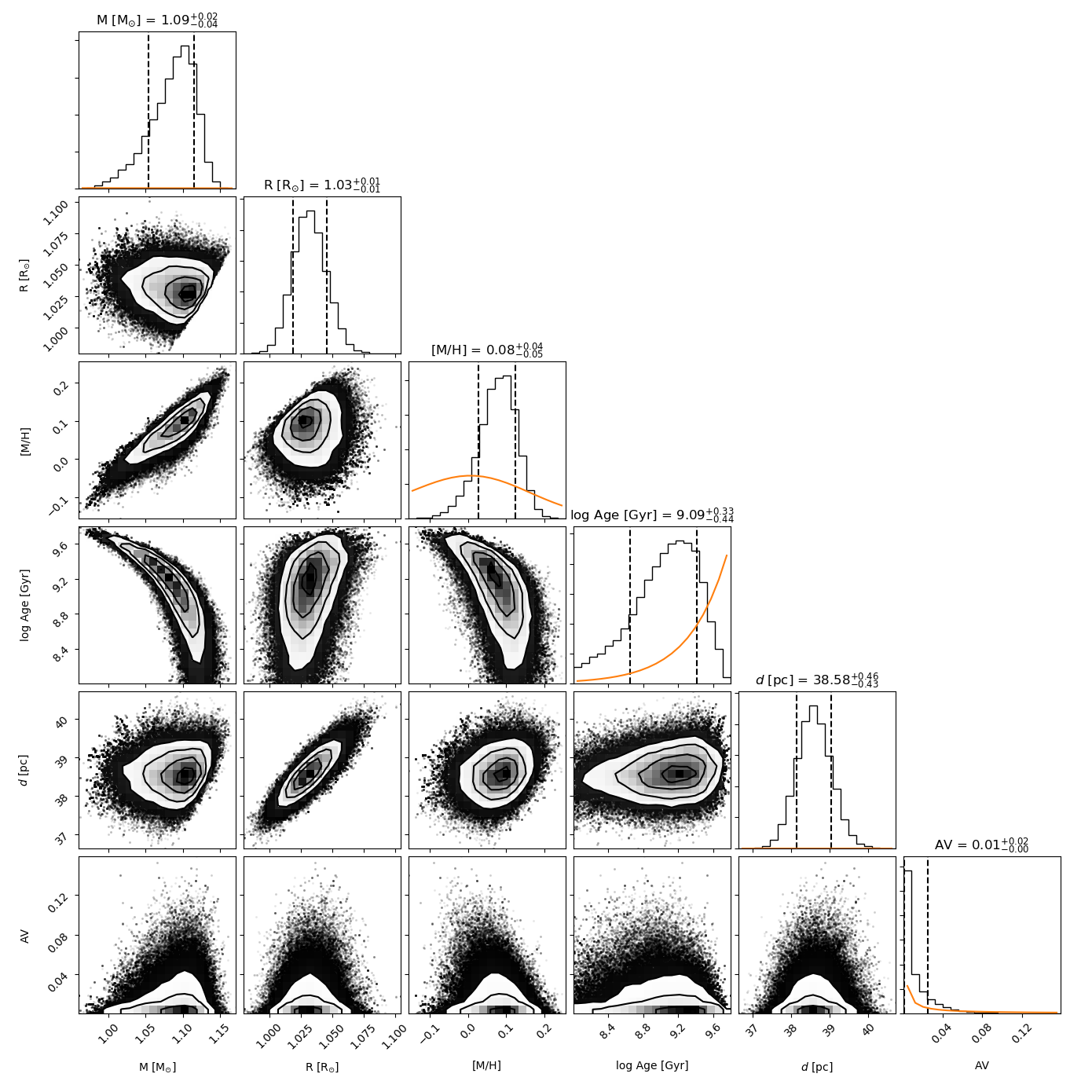}
\caption{Corner plot showing marginalized posterior probability distributions for HD 105. Orange lines indicate the priors. Note, radius is not a model parameter. \label{fig:isochrones_out}}
\end{figure*}

For each star, we used the \texttt{emcee} python module \citep{Foreman-Mackey2013} to sample the posterior with an affine-invariant ensemble sampler \citep{Goodman2010}. We used 100 walkers with 2000 burn-in steps and 5000 sampling steps. We initialized the walkers based on parameter estimates that maximized the posterior. We remove chains with an acceptance fraction $<$ 0.1 and exclude from further analysis any star whose maximum integrate autocorrelation time is greater than 1/3 the number of sampling steps. Figures~\ref{fig:isochrones_in}~\&~\ref{fig:isochrones_out} show corner plots for the modeled observables and model parameters, respectively, for one representative star, HD 105. The input observables are well reproduced by the models to within measurement uncertainties. We take the median of the marginalized age posterior as the best fit age.

\end{document}